# Machine-Learning-Based Interatomic Potentials for Group IIB to VIA Semiconductors: Towards a Universal Model

Jianchuan Liu [a, †], Xingchen Zhang [b, †], Tao Chen [c], Yuzhi Zhang [d], Duo Zhang [d, e], Linfeng Zhang [d, f], and Mohan Chen [*,b,f]

[a.] *School of Electrical Engineering and Electronic Information, Xihua University, Chengdu, 610039, P. R. China*

[b.] *HEDPS, CAPT, College of Engineering, Peking University, Beijing, 100871, P. R. China*

[c] *HEDPS, CAPT, School of Physics, Peking University, Beijing, 100871, P. R. China*

[d.] *DP Technology, Beijing 100080, P. R. China*

[e.] *Center for Machine Learning Research, Academy for Advanced Interdisciplinary Studies, Peking University, Beijing, 100871, P. R. China*

[f.] *AI for Science Institute, Beijing 100080, P. R. China*

∗ *Corresponding author. Email: mohanchen@pku.edu.cn*

## Abstract

Rapid advancements in machine-learning methods have led to the emergence of machine-learning-based interatomic potentials as a new cutting-edge tool for simulating large systems with *ab initio* accuracy. Still, the community awaits universal inter-atomic models that can be applied to a wide range of materials without tuning neural network parameters. We develop a unified deep-learning inter-atomic potential (the DPA-Semi model) for 19 semiconductors ranging from group IIB to VIA, including Si, Ge, SiC, BAs, BN, AlN, AlP, AlAs, InP, InAs, InSb, GaN, GaP, GaAs, CdTe, InTe, CdSe, ZnS, and CdS. In addition, independent deep potential models for each semiconductor are prepared for detailed comparison. The training data are obtained by performing density functional theory calculations with numerical atomic orbitals basis sets to reduce the computational costs. We systematically compare various properties of the solid and liquid phases of semiconductors between different machine-learning models. We conclude that the DPA-Semi model achieves GGA exchange-correlation functional quality accuracy and can be regarded as a pre-trained model towards a universal model to study group IIB to VIA semiconductors.

## 1. Introduction

Semiconductor materials play a crucial role in the development of modern society. In particular, the IIB to VIA group semiconductors, which are compound semiconductors composed of elements from groups IIB to VIA of the periodic table, possess excellent optoelectronic properties and are widely used in photovoltaic, optoelectronics, thermoelectrics, and other energy conversion fields.[1, 2, 3, 4] For example, silicon carbide (SiC) has found widespread industrial applications because of its excellent wear resistance, corrosion resistance, elevated temperature strength, as well as its high thermal conductivity and wide band gap.[5, 6, 7, 8] Boron arsenide (BAs) was initially synthesized in 1958 [9] but was recently confirmed to possess high charge carrier mobility and thermal conductivity.[10, 11, 12] Therefore, there is a promising prospect of utilizing this material to alleviate the current bottleneck issue in chip cooling. On the other hand, state-of-the-art simulation tools can complement experiments by elucidating experimental phenomena or predicting experimental outcomes, thereby providing invaluable information or better design principles.

Among the simulation tools, the atomistic-level simulation tools can describe interactions of semiconductors from a microscopic perspective, providing valuable insights into the fundamental processes governing the behavior of semiconductor materials. In particular, quantum-mechanics-based first-principles methods are able to predict the properties of semiconductors without reliance on experimental data. Among them, density functional theory (DFT)[13, 14] is one of the most widely used methods that can predict various properties of semiconductor materials.[15, 16, 17] Taking SiC as an example, it exists in various polytypes such as the 3C, 2H, 4H, and 6H structures, etc. Among them, the 3C structure is a cubic crystal lattice, while the 2H, 4H, and 6H polytypes refer to hexagonal crystal lattices with different stacking sequences of Si-C bilayers[18, 19, 20, 21, 22, 23, 24, 25, 26]. The relationship between atomic structures and mechanical properties in single crystal and polycrystalline solid phases,[22, 27, 28] as well as the stability and mobility of screw dislocations in 3C, 2H, and 4H-SiC can be calculated by DFT.[8] In addition, the compression mechanical properties[25] and the elastic properties[26] of 3C, 4H, and 6H-SiC at ambient and high pressure, and the electronic structures and optical properties of vacancy-doped 3C-SiC systems were also studied by DFT.[29]

However, simulations of semiconductor materials for industrial applications, which often demand a large number of atoms (typically millions or more), remain a formidable challenge when using the DFT method. Recently, rapid advancements in machine-learning methods have led to the emergence of machine-learning-based interatomic potentials. Among them, the deep potential (DP) model[30] with first-principles accuracy[31, 32, 33, 34, 35, 36] has emerged as a cutting-edge tool for simulating large-size systems[37] and long-timescale properties across a wide range of materials. Taking semiconductor materials as examples, all of the following works adopted the DP models. The thermal conductivity for the crystalline, liquid, and amorphous phases of Si,[38] as well as the thermal conductivity and phonon transport properties of $\beta$-$Ga_2O_3$ were accurately predicted.[39] In addition, the temperature-dependent microwave dielectric permittivity of $\beta$-$Ga_2O_3$ was calculated.[40] For the SiC materials, the thermal transport and mechanical properties were systematically investigated,[41] and the infrared resonance frequency and phonon linewidth were accurately predicted.[20] The DP models were also utilized in research related to heat transfer of semiconductor interfaces such as the Si/Ge interface.[42] These research outcomes demonstrate that the machine-learning-based DP method plays a key role in atomistic simulations of semiconductor materials.

However, when addressing similar properties of a wide range of materials, generating a DP model for each material is not only computationally demanding but also considerably time-consuming. In addition, some inconsistencies may exist among the models due to the different training samples used. In this regard, there is a strong demand in the community for generating a universal model that can simulate a wide range of materials. In 2022, a deep potential model with an attention mechanism was proposed by Zhang et al.,[43] i.e., the DPA-1 model. The model was trained on a large number of atomic datasets in terms of a variety of elements and showed satisfactory accuracy. Nowadays, the DPA-1 model can be readily applied to study real scientific issues with a small amount of additional effort.

Machine learning interatomic potential functions such as the DP and DPA-1 models are typically trained on a large number of DFT datasets. Typically, DFT calculations are performed with the Plane-Wave (PW) basis set.[27] The PW basis set has several advantages. For instance, the accuracy of DFT calculations can be controlled by a single value of energy cutoff. In addition, the PW basis sets are orthogonal, and no Pulay forces need to be evaluated. Notably, the PW basis set exhibits unfavorable scaling when the system size is large, typically around a few hundred atoms, resulting in large computational costs for generating training data. In this work, to reduce the computational costs of DFT, numerical atomic orbitals (NAO)[44, 45, 46, 47, 48] as a basis set are adopted to solve the Kohn-Sham equation. Specifically, we utilize the Atomic-orbital Based Ab-initio Computation at UStc (ABACUS) package[49], which supports the NAO basis set.[44, 45] The computational costs are relatively smaller than the PW basis set. The NAO basis set has been used in several applications and is suitable for studying large systems.[50, 51, 52, 53]

In this work, we generated first-principles data for 19 bulk semiconductors ranging from group IIB to VIA, namely, Si, Ge, SiC, BAs, BN, AlN, AlP, AlAs, InP, InAs, InSb, GaN, GaP, GaAs, CdTe, InTe, CdSe, ZnS, CdS. We used the ABACUS 3.2 package based on the numerical atomic orbitals basis set with the Perdew-Burke-Ernzerhof (PBE)[54] exchange-correlation functional generated atomic datasets to reduce the production cost of the data (Sec. 2.1). We adopted the atomic datasets as training data to generate an attention-based deep potential model using the DPA-1 method, which we named as the DPA-Semi model (Sec. 2.2). The procedures to generate the DPA-Semi model are shown in Fig. 1. For comparison, we also generated DP models for each of the 19 semiconductors mentioned above (Sec. 2.2). The calculated lattice constants, bulk moduli, shear moduli and Young's moduli of different semiconductors indicate that the results of DPA-Semi model are consistent with the values of our DFT calculation. (Sec. 3.1). Furthermore, the DPA-Semi model was utilized to investigate the liquid and amorphous structures (Sec. 3.2) and melting temperatures (Sec. 3.3) of various semiconductor materials, and the results are consistent with experiments and other calculations.

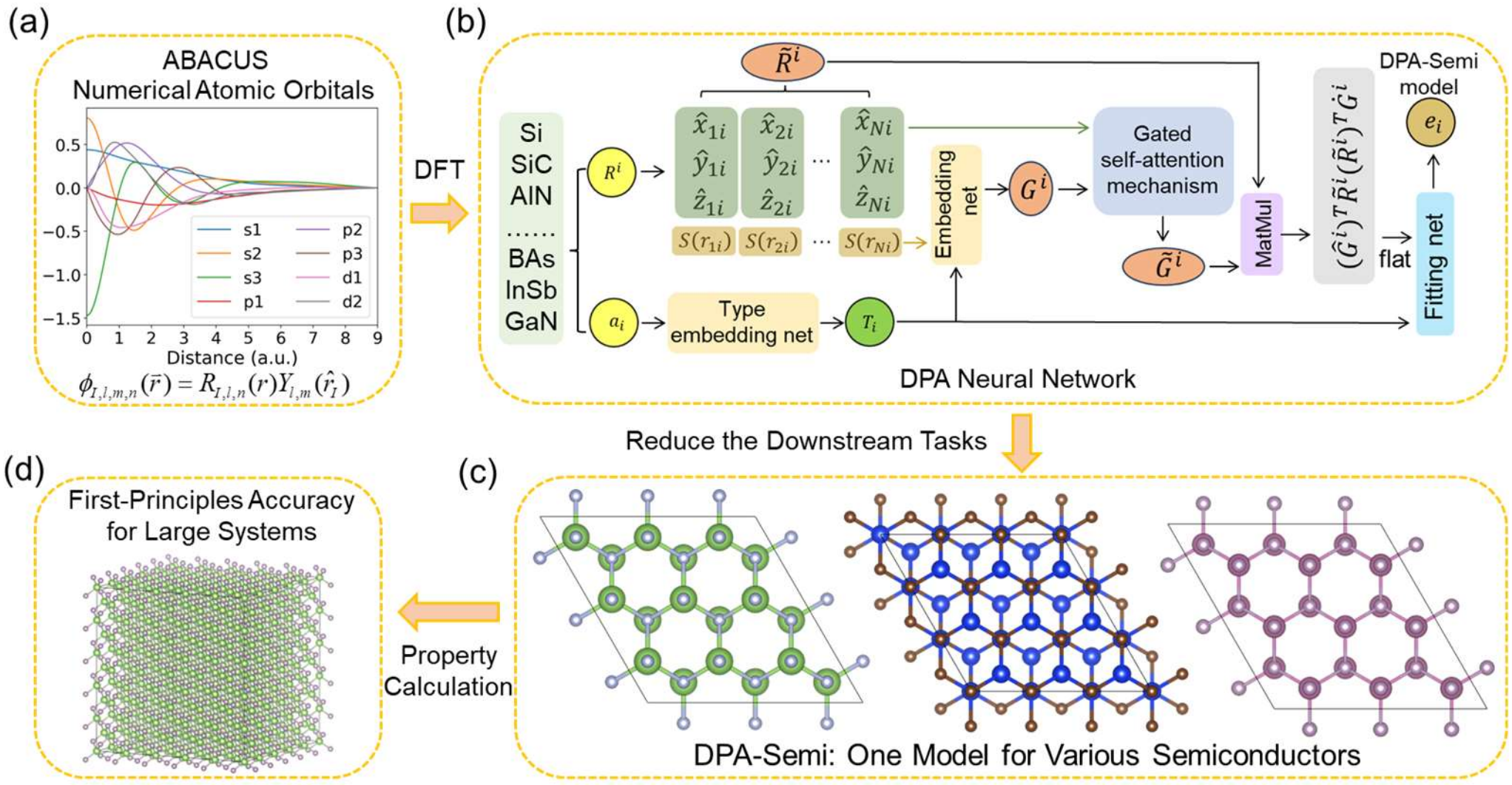

**Fig. 1** Procedures for developing the DPA-Semi model. (a) Generate atomic datasets using the ABACUS package based on the numerical atomic orbitals as basis set; (b) Generate the DPA-Semi model via the Gated self-

attention mechanism based on the DFT atomic datasets; (c) The DPA-Semi model can be used for various kinds of semiconductors, and reduce the computational costs of downstream tasks; (d) The DPA-Semi model is readily applied to calculate properties of large-systems with GGA quality accuracy.

## 2. Methods

### 2.1 Density Functional Theory

All of the DFT calculations were performed with the ABACUS 3.2 package.[49] The generalized gradient approximation (GGA) in the form of the Perdew-Burke-Ernzerhof (PBE)[54] was used for the exchange-correlation functional. The norm-conserving pseudopotentials[55, 56] were employed, and the valence configurations for the elements were [B]$2s^2 2p^1$, [C]$2s^2 2p^2$, [N]$2s^2 2p^3$, [Al]$3s^2 3p^1$, [Si]$3s^2 3p^2$, [P]$3s^2 3p^3$, [S]$3s^2 3p^4$, [Zn]$3s^2 3p^6 3d^{10} 4s^2$, [Ga]$4s^2 4p^1$, [Ge]$4s^2 4p^2$, [As]$4s^2 4p^3$, [Cd]$4s^2 4p^6 4d^{10} 5s^2$, [In]$4d^{10} 5s^2 5p^1$, [Sb]$5s^2 5p^3$, and [Te]$4d^{10} 5s^2 5p^4$, respectively. The energy cut-off was set to 100 Ry. The Monkhorst-Pack[57] *k*-points mesh was set with the spacing being 0.08 $Bohr^{-1}$. We employed the Gaussian smearing method with a smearing width of 0.002 Ry. The electronic iteration convergence threshold was set to $10^{-6}$.

The triple-zeta plus double polarization (TZDP) numerical atomic orbitals basis sets were used for all of the DFT calculations. The numerical atomic orbitals were chosen to be 3s3p2d for B, 3s3p2d for C, 3s3p2d for N, 3s3p2d for Al, 3s3p2d for Si, 3s3p2d for P, 3s3p2d for S, 6s3p3d2f for Zn, s3p2d for Ga, 3s3p2d for Ge, 3s3p2d for As, 6s3p3d2f for Cd, 3s3p3d2f for In, 3s3p2d for Sb, and 3s3p3d2f for Te, respectively. The cutoffs of numerical atomic orbitals were all set to 9 a.u.

### 2.2 Deep Potential Generation

In this work, we constructed machine-learning-based models for 19 kinds of semiconductors ranging from group IIB to VIA, namely, Si, Ge, SiC, BAs, BN, AlN, AlP, AlAs, InP, InAs, InSb, GaN, GaP, GaAs, CdTe, InTe, CdSe, ZnS, CdS. For each semiconductor, we first selected a variety of crystal structures, as detailed in Table 1. Next, random perturbations were performed on the atomic coordinates by adding values drawn from a uniform distribution in the range of [-0.01, 0.01]. We also changed the cell vectors by a symmetric deformation matrix constructed by adding random noise drawn from a uniform distribution in the range of [-0.03, 0.03]. Five steps of AIMD simulations were performed for all the perturbed structures to produce labeled data with energies, forces, and virial tensors calculated from DFT. These labeled data were used to form the initial data sets.

**Table 1.** Crystal structures adopted for generating first-principles data for 19 semiconductors, as well as the ranges of temperatures used for generating the corresponding machine-learning-based models. The pressure range is set to 0-5 GPa.

| System | Temperature (K) | Crystal Structures |
|---|---|---|
| Si | 0- 3200 | Cubic ($Fd\bar{3}m$), Hexagonal ($P6_3/mmc$), Hexagonal (P6/mmm), Tetragonal ($I4_1/amd$) |
| Ge | 0-2600 | Cubic ($Fd\bar{3}m$), Hexagonal ($P6_3/mmc$), Tetragonal ($I4_1/amd$) |
| SiC | 0-5950 | Cubic ($F\bar{4}3m$), Hexagonal ($P6_3mc$) |
| BAs | 0-4600 | Cubic ($F\bar{4}3m$), Hexagonal ($P6_3mc$) |
| BN | 0-6600 | Cubic ($F\bar{4}3m$), Hexagonal ($P6_3mmc$) |
| AlN | 0-7200 | Cubic ($F\bar{4}3m$), Hexagonal ($P6_3mc$), Hexagonal ($P6_3/mmc$) |
| AlP | 0-4600 | Cubic ($F\bar{4}3m$), Hexagonal ($P6_3mc$) |
| AlAs | 0-4200 | Cubic ($F\bar{4}3m$), Hexagonal ($P6_3/mmc$) |
| InP | 0-3200 | Cubic ($F\bar{4}3m$), Hexagonal ($P6_3mc$) |
| InAs | 0-2600 | Cubic ($F\bar{4}3m$), Hexagonal ($P6_3mc$) |
| InSb | 0-2200 | Cubic ($F\bar{4}3m$), Hexagonal ($P6_3mc$) |
| GaN | 0-4000 | Cubic ($F\bar{4}3m$), Hexagonal ($P6_3mc$) |
| GaP | 0-3500 | Cubic ($F\bar{4}3m$), Hexagonal ($P6_3mc$) |
| GaAs | 0-3000 | Cubic ($F\bar{4}3m$), Hexagonal ($P6_3mc$) |
| CdTe | 0-2650 | Cubic ($F\bar{4}3m$), Hexagonal ($P6_3mc$) |
| InTe | 0-1950 | Cubic ($Fm\bar{3}m$), Cubic ($Pm\bar{3}m$) |
| CdSe | 0-3250 | Cubic ($F\bar{4}3m$), Hexagonal ($P6_3mc$) |
| ZnS | 0-3950 | Cubic ($F\bar{4}3m$) |
| CdS | 0-4050 | Cubic ($F\bar{4}3m$), Hexagonal ($P6_3mc$) |

Next, we utilized the Deep Potential Generator (DP-GEN 0.11.1 package)[58] to generate Deep Potential (DP) models[30] for each semiconductor. The initial data were trained by the DeePMD-kit 2.2.8 package.[59] We adopted three hidden layers for the embedding network for the DP models, with sizes of 25, 50, and 100. In addition, three hidden layers with sizes of 240, 240, and 240 were selected for the fitting network. An exponentially decaying learning rate was chosen to change from $1.0\times10^{-3}$ to $3.5\times10^{-8}$. During the optimization process, the prefactor of the energy (force) term in the loss function changes from 0.02 to 1 (1000 to 1). The DP model was trained for $4.0\times10^{5}$ steps with the cutoff radius being 8 Å. Four DP models were generated for each training process, where the same reference dataset was used, but the initial parameters for the deep neural network were different.

Finally, we performed MD simulations with the DP model with temperatures ranging from 0 K to twice the melting temperatures of each semiconductor (detailed temperature settings are shown in Table 1) and pressures ranging from 0 to 5 GPa to explore new configurations using the LAMMPS (23 Jun 2022) package.[60] We did not include perturbations in the initial configurations. More details regarding the adopted crystal structures, as well as the selected ranges of temperatures and pressures for exploring the configuration space in each iteration, are shown in Table 1. During each iteration, a maximum of 60 candidate configurations were selected for each semiconductor crystal structure. These configurations were added to the training set for the next iteration after calculating their energies, atomic forces, and virial tensor using the DFT method. All iterations were done automatically with the DP-GEN 0.11.1 package. The iterations continued until the proportion of candidate configurations was less than 5% and remained almost unaltered for another few iterations. After the DP-GEN iterations were converged, we trained the collected data for $1.2 \times 10^7$ steps with "se_e2_a" descriptors (DP model)[59, 61] and "se_atten" descriptors (DPA-Semi model)[43] using DeePMD-kit 2.2.8 package, respectively. The number of data sets for 19 semiconductors is shown in Table 2. In addition, the AlN system has the largest amount of data sets with 29715 frames. Because its melting temperatures is high, the iterative temperature reaches twice its melting temperatures (Table 1), so the amount of data sets sampled is relatively large. The InSb system has the least amount of data sets, which is 2415 frames. This is caused by the fact that the melting temperatures of InSb is relatively small, indicating that the amount of data sampled is relatively small.

Furthermore, Table 2 shows the root-mean-square errors (RMSEs) of the energies and forces predicted by the DP and DPA-Semi models. Generally, smaller RMSEs of energy and forces imply a relatively higher accuracy of the machine learning models. We find the smallest RMSEs of energy (0.005 eV/atom) and forces (0.11 eV/Å) are found for the Ge system by using the DP model. For the DPA-Semi model, the energy RMSEs of the BAs system are the smallest, which is 0.0032 eV/atom, and the force RMSEs of the InSb system are the smallest, which is 0.11 eV/Å. On the contrary, for the DP model, the SiC system has the largest RMSEs of energy (0.012 eV/atom) and forces (0.37 eV/Å). For the DPA-Semi model, the largest RMSEs of energy (0.010 eV/atom) and force (0.37 eV/Å) come from the ZnS system and the BN system, respectively. The lattice constant and elastic constants computed by the DP and DPA models agree well with the DFT results. The data sets and the deep-learning potential models (DP and DPA-Semi) are available for download.[62, 63] In addition, examples of the DP and DPA-Semi models can be run by Bohrium Notebook.[64]

**Table 2.** Root-mean-square errors (RMSEs) of the total energy (meV/atom) and forces (meV/Å) predicted by each DP model and the DPA-Semi model, and the number of data sets generated for 19 semiconductors after the DP-GEN iterations. Note that the data to calculate RMSEs are from both solid and liquid phases.

| | Si | Ge | SiC | BAs | BN | AlN | AlP | AlAs | InP | InAs |
|---|---|---|---|---|---|---|---|---|---|---|
| **Energy (DP)** | 5.53 | 3.90 | 11.61 | 4.53 | 10.49 | 11.48 | 9.11 | 7.92 | 7.95 | 6.42 |
| **Energy (DPA)** | 6.63 | 5.42 | 9.04 | 3.24 | 6.11 | 10.18 | 8.74 | 7.81 | 8.97 | 7.37 |
| **Force (DP)** | 120.27 | 114.61 | 370.00 | 151.34 | 354.74 | 324.40 | 216.74 | 187.28 | 191.00 | 152.87 |
| **Force (DPA)** | 181.22 | 120.06 | 303.34 | 142.37 | 367.91 | 283.25 | 176.01 | 163.56 | 182.47 | 139.87 |
| **Frame** | 16923 | 18966 | 12749 | 4236 | 18045 | 29715 | 6988 | 5043 | 8601 | 7477 |
| | **InSb** | **GaN** | **GaP** | **GaAs** | **CdTe** | **InTe** | **CdSe** | **ZnS** | **CdS** | |
| **Energy (DP)** | 4.29 | 9.71 | 8.46 | 6.70 | 7.33 | 6.20 | 5.88 | 8.61 | 6.77 | |
| **Energy (DPA)** | 6.12 | 4.98 | 6.45 | 6.53 | 7.94 | 7.75 | 6.94 | 10.36 | 7.24 | |
| **Force (DP)** | 112.88 | 263.22 | 211.88 | 166.11 | 145.47 | 133.63 | 127.43 | 182.99 | 151.63 | |
| **Force (DPA)** | 105.53 | 311.29 | 192.74 | 152.66 | 138.55 | 132.70 | 127.12 | 192.79 | 145.64 | |
| **Frame** | 2415 | 9005 | 11950 | 7045 | 7753 | 7375 | 11164 | 8763 | 12589 | |

### 2.3 Elastic constants

Elastic constants ($C_{\mathrm{ij}}$) can be determined by performing a linear least-squares fit between stress and strain for a series of small deformations in the crystal lattice[65], following Hooke's law:

$$\sigma_{ij} = C_{ijkl} \cdot \epsilon_{kl} \qquad (1)$$

where $\sigma_{ij}$ is the stress tensor, $\epsilon_{kl}$ is the strain tensor, and $C_{ijkl}$ is the fourth-rank elastic stiffness tensor, which can be further simplified using the two-indices Voigt notation as a 6×6 $C_{ij}$ matrix.

In this paper, the zero-temperature elastic constants ($C_{\mathrm{ij}}$) are evaluated by the DP-GEN software package[58], which is based on the Python Materials Genomics (pymatgen) library[65, 66]. The bulk modulus (B), shear modulus (G), and Young's modulus (E) are estimated using the corresponding equations based on the elastic constants[67, 68]. For a cubic crystal structure, these properties are computed via the formulas of

$$\boldsymbol{B} = \frac{1}{3}(C_{11} + 2C_{12}),\ \boldsymbol{G} = \frac{1}{5}(C_{11} - C_{12} + 3C_{44}),\ \boldsymbol{E} = \frac{9B}{1+\frac{3B}{G}}. \qquad (2)$$

For a hexagonal crystal structure, the formulas are

$$\boldsymbol{B} = \frac{1}{9}(2(C_{11} + C_{12}) + 4C_{13} + C_{33}),$$

$$\boldsymbol{G} = \frac{1}{30}(C_{11} + C_{12} + 2C_{33} - 4C_{13} + 12C_{44} + 12C_{66}),\ \boldsymbol{E} = \frac{9B}{1+\frac{3B}{G}} \qquad (3)$$

### 2.4 Mean square displacements

The Mean square displacement (MSD) is defined as $u^2(t)$ according to

$$u^2(t) = \frac{1}{N}\sum_{i=1}^{N}[R_i(t) - R_i(0)]^2 \qquad (4)$$

where $R_i(t)$ is the atom position of atom *i* after t time of simulation. *N* is the total number of atoms. MSD provides a method to observe whether the system is in a solid or liquid state. If the system is in a solid state, then MSD oscillates around a constant value. This indicates that all atoms are confined to specific positions. However, if some atoms melt, the MSD linearly increases. Utilizing the DP model and DPA-Semi model, we calculated the melting temperature data of semiconductor materials based on the observation of variations in MSD.

### 2.5 Radial distribution functions

The radial distribution function g(r) describes the local structure of atoms and is defined as:

$$g(r) = \frac{V}{4\pi r^2 N^2}\langle\sum_{i=1}^{N}\sum_{j=1,j\neq i}^{N}\delta(r - |\boldsymbol{r}_i - \boldsymbol{r}_j|)\rangle \ , \qquad (5)$$

where *V* is the cell volume, *N* is the number of atoms, $\boldsymbol{r}_i$ and $\boldsymbol{r}_j$ are atomic coordinates of atoms *i* and *j*, and $\langle\cdots\rangle$ means the time or ensemble average. In this manuscript, we computed the Radial Distribution Function (RDF) for amorphous silicon, liquid silicon, and amorphous InSb systems utilizing the DPA-Semi model. A comparative analysis with experimental data was undertaken to substantiate the accuracy of the DPA-Semi model.

## 3. Results and Discussion

### 3.1 Lattice constants and mechanical properties

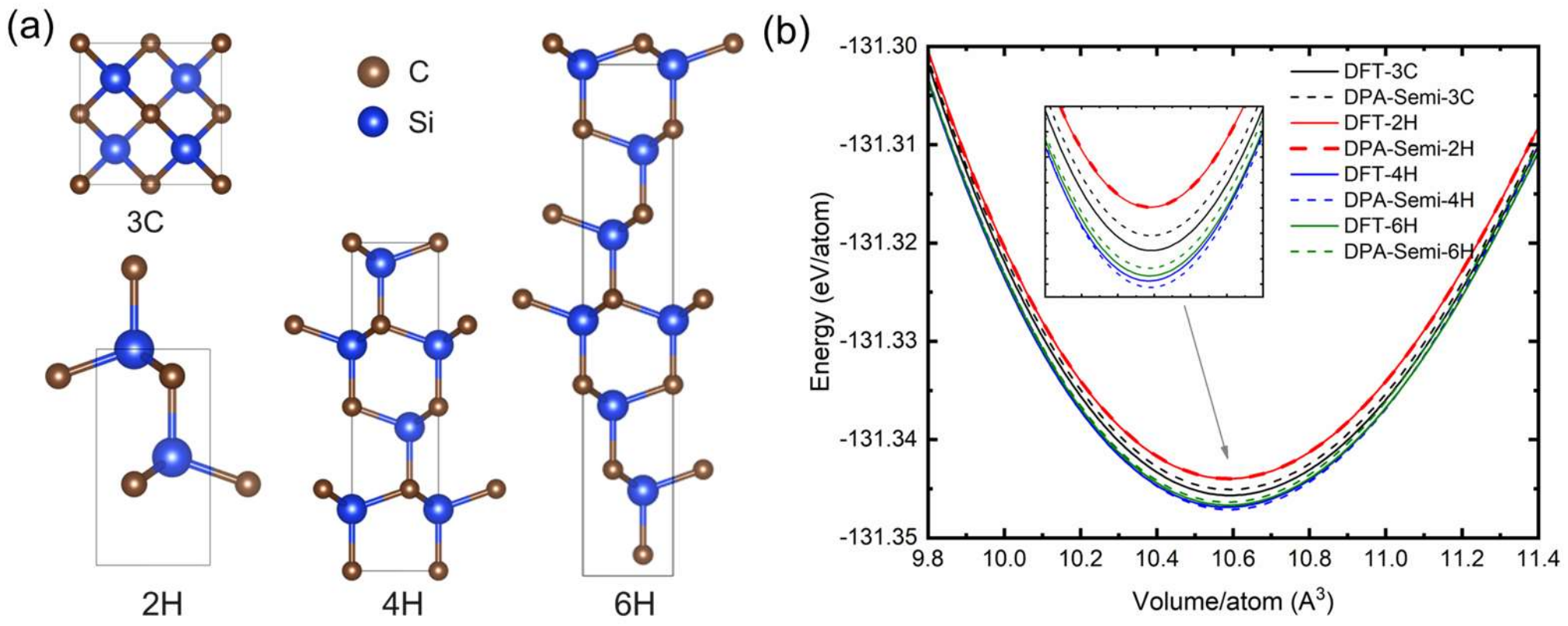

**Fig.2** (a) Crystal structures of the 3C-, 2H-, 4H-, and 6H-SiC polytypes. The blue and gray spheres represent the Si and C atoms, respectively. (b) Curves of energy vs. average atomic volume for 3C-, 2H-, 4H-, and 6H-SiC polytypes calculated by the DFT (solid lines) and DPA-Semi (dashed lines) methods.

Fig. 2(a) illustrates the stable structures of silicon carbide (SiC), which include the 3C, 2H, 4H, and 6H structures with close energies[18, 19, 20, 21, 22, 23, 24, 25, 26]. Therefore, it is an ideal example for testing the accuracy of the DPA-Semi model. In detail, the SiC structures are based on different stacking patterns in the cubic and hexagonal diamond structures. As shown in Fig. 2(a), the 3C structure is a zinc blende structure with the ABC stacking, while the 2H, 4H, and 6H structures take the wurtzite structure and have the AB, ABCB, and ABCACB stackings.[18, 19] In this regard, the 3C-, 2H-, 4H-, and 6H-SiC structures have 8, 4, 8, and 12 atoms per primitive unit cell, respectively. We find the cells shown in Fig. 2(a) become more elongated along the cross-plane direction of *n*H-SiC with increasing *n*.

Fig. 2(b) shows the computed total energy with respect to the volume per atom for the 3C-, 2H-, 4H-, and 6H-SiC polytypes calculated by the DFT and DPA-Semi methods. We find the calculated total energies satisfy the inequality relation: $E_{2H} > E_{3C} > E_{6H} > E_{4H}$. This data indicates that the 4H-SiC structure is more stable than the other three structures, which agrees with previous DFT results.[25, 26, 69] Although the energy differences between the polytypes are small, on the order of meV/atom, both DFT and DPA-Semi models correctly distinguish these structural energy differences, demonstrating that the DPA-Semi model owns a satisfactory GGA quality accuracy.

Table 3 lists the in- and cross-plane lattice constants (*a* and *c*) and elastic constants ($C_{11}$, $C_{12}$, $C_{13}$, $C_{33}$, $C_{44}$, and $C_{66}$) for the 3C-, 2H-, 4H- and 6H-SiC structures. The data are obtained by ABACUS[49] with the NAO basis and the PBE functional, the DP models,[30] the DPA-Semi mode,[43] DFT data from references,[70,71,72] and available experimental data. Our results for the lattice constants and elastic constants of the 3C-, 2H-, 4H-, and 6H-SiC structures agree well with other calculated results[8, 25, 26, 73] and experimental data from Refs. [74, 75, 76, 77].

Taking the 3C-SiC structure as an example. Table 3 lists the elastic constants ($C_{11}$, $C_{12}$, and $C_{44}$) of 3C-SiC. We find the $C_{11}$ values obtained from the DP models, the DPA-Semi model, the ABACUS package, and the experiments are 385, 379, 379, and 395 GPa, respectively. In this regard, the maximum deviation between our calculations and the experimental results is 4.1 %, which suggests that the DP models and the DPA-Semi model are sufficiently accurate. On the other hand, we compare the deviation of elastic constants ($C_{11}$, $C_{12}$, and $C_{44}$) between DP and DFT calculations. In particular, we observe that the discrepancies for the $C_{12}$ value of the 3C, 2H, 4H, and 6H-SiC structures between DP and DFT are 3.3%, 4.4%, 9.4%, and 10.8%, respectively. On the other hand, we also compare the discrepancies for the $C_{12}$ value of the 3C, 2H, 4H, and 6H-SiC structures between DPA-Semi and DFT are 0.8%, 3.1%, 5.4%, and 8.8%, respectively. It indicates that the DPA-Semi model predicts slightly closer values of the $C_{12}$ elastic constants than the DP model when compared to the DFT results.

Table 3 also shows the bulk modulus (B), shear modulus (G), and Young's modulus (E) of the 3C-, 2H-, 4H-, and 6H-SiC structures. As listed in Table 3, the DP and DPA-Semi models reproduce the elastic moduli of the SiC structures with DFT accuracy. For example, the maximum deviation of bulk modulus for 3C-, 2H-, 4H-, and 6H-SiC obtained from the DPA-Semi model and the DFT calculations is 1.9% (2H-SiC). These results demonstrate that the DP and DPA-Semi models are suitable for studying the mechanical properties of SiC polytypes. Besides the results for SiC, Table 4 shows the lattice constants and elastic moduli of 19 semiconductors from group IIB to VIA. We find that the results obtained from the DP and DPA-Semi models are in excellent agreement with the DFT results calculated by ABACUS. These results offer reliable evidence that the DP and DPA-Semi models can be employed to study the physical mechanisms of group IIB to VIA semiconductor systems with GGA quality accuracy. The DP/DPA-Semi results of bulk modulus are in excellent agreement with other DFT results.[78, 27, 28] Taking GaN, GaP, and GaAs as examples, the maximum deviation of the bulk modulus between DP/DPA-Semi and DFT is 1.1 % for GaP. From Table 4, the *B* of GaN, GaP, and GaAs is 156, 76, and 61 GPa calculated by the DPA-Semi model, respectively. If compared to the experimental bulk modulus from Ref. [79], i.e., the *B* of GaN, GaP, and GaAs is 210, 89, and 76 GPa, respectively, the calculated values for the bulk modulus which were obtained in our calculations or other DFT results[78] are underestimated due to the PBE functional failure to account for non-local correlation effects among electrons in a material.[78] This effect is particularly substantial for materials with metallic or covalent properties where electron-electron interactions are strong.

To facilitate comparison, Fig. 3 shows the comparison between ABACUS, DP, DPA-Semi, and experimental data in terms of lattice constants and bulk modulus. We find that the maximum deviation of lattice constants between the computational methods and experimental data is the CdSe system (26.9%). The maximum deviation of bulk modulus between ABACUS/DP/DPA-Semi and experimental data is the InTe system (44.9%).

**Table 3.** Lattice constant (*a* and *c*, in Å), elastic constants $C_{ij}$ (in GPa), bulk modulus (*B*, in GPa), shear modulus (*G*, in GPa), and Young's modulus (*E*, in GPa) for the 3C-, 2H-, 4H-, and 6H-SiC polytypes structures calculated by the ABACUS (using NAO basis set with the PBE functional, labeled as NAO-PBE) package, the DP model, the DPA-Semi model, the PW-PBE method (using PW basis set with the PBE functional), and the PW-LDA method. Experimental results are also presented as comparisons.

| **Methods** | ***a*** | ***c*** | $C_{11}$ | $C_{12}$ | $C_{13}$ | $C_{33}$ | $C_{44}$ | $C_{66}$ | ***B*** | ***G*** | ***E*** |
|---|---|---|---|---|---|---|---|---|---|---|---|
| ***3C*** | | | | | | | | | | | |
| NAO-PBE | 4.392 | - | 379 | 125 | - | - | 239 | - | 210 | 194 | 445 |
| DP | 4.392 | - | 385 | 121 | - | - | 242 | - | 212 | 195 | 448 |
| DPA-Semi | 4.391 | - | 373 | 129 | - | - | 238 | - | 210 | 192 | 445 |
| PW-PBE[a] | 4.366 | - | 376 | 129 | - | - | 246 | - | 211 | 176 | 310 |
| PW-PBE[b] | 4.380 | - | 382 | 128 | - | - | 239 | - | - | - | - |
| PW-LDA[c] | 4.328 | - | 369 | 138 | - | - | 226 | - | 215 | 181 | 425 |
| Exp. | 4.360[e] | - | 395[m] | 123[m] | - | - | 236[m] | - | 225[g] | 192[h] | 469[i] |
| ***2H*** | | | | | | | | | | | |
| NAO-PBE | 3.100 | 5.087 | 490 | 96 | 47 | 529 | 150 | 196 | 210 | 187 | 433 |
| DP | 3.100 | 5.086 | 484 | 100 | 50 | 529 | 151 | 191 | 211 | 185 | 430 |
| DPA-Semi | 3.100 | 5.082 | 497 | 101 | 55 | 526 | 158 | 197 | 214 | 190 | 440 |
| PW-PBE[d] | 3.079 | 5.053 | 506 | 92 | 46 | 542 | 154 | - | 213 | 191 | 441 |
| PW-PBE[b] | 3.088 | 5.083 | 490 | 93 | 52 | 533 | 153 | - | - | - | - |
| Exp. | 3.079[f] | 5.053[f] | - | - | - | - | - | - | - | - | - |
| ***4H*** | | | | | | | | | | | |
| NAO-PBE | 3.102 | 10.156 | 490 | 93 | 49 | 528 | 156 | 194 | 210 | 189 | 437 |
| DP | 3.102 | 10.162 | 490 | 102 | 57 | 551 | 162 | 193 | 218 | 191 | 444 |
| DPA-Semi | 3.103 | 10.158 | 483 | 105 | 49 | 529 | 159 | 189 | 211 | 189 | 437 |
| PW-PBE[a] | 3.084 | 10.096 | 497 | 97 | 49 | 529 | 154 | 199 | 213 | 184 | 475 |
| PW-PBE[b] | 3.090 | 10.178 | 498 | 91 | 52 | 535 | 153 | - | - | - | - |
| PW-LDA[c] | 3.067 | 10.068 | 379 | 116 | - | - | 242 | - | 204 | 197 | 448 |
| Exp. | 3.073[e] | 10.052[e] | 501[j] | 111[j] | 52[j] | 553[j] | 163[j] | - | 215[k] | 131[k] | 450[l] |
| ***6H*** | | | | | | | | | | | |
| NAO-PBE | 3.103 | 15.226 | 491 | 91 | 49 | 528 | 159 | 194 | 210 | 190 | 439 |
| DP | 3.103 | 15.227 | 486 | 102 | 54 | 532 | 164 | 191 | 214 | 190 | 440 |
| DPA-Semi | 3.103 | 15.227 | 479 | 106 | 49 | 527 | 162 | 190 | 211 | 190 | 436 |
| PW-PBE[a] | 3.085 | 15.138 | 493 | 100 | 53 | 532 | 156 | 196 | 214 | 184 | 469 |
| PW-LDA[c] | 3.074 | 15.100 | 376 | 118 | - | - | 238 | - | 204 | 194 | 442 |
| Exp. | 3.081[e] | 15.120[e] | 501[j] | 111[j] | 52[j] | 553[j] | 163[j] | - | 215[k] | 131[k] | 450[l] |

[a] Ref.[25], [b] Ref.[8], [c] Ref.[26], [d] Ref.[73], [e] Ref.[74], [f] Ref.[76], [g] Ref.[80], [h] Ref.[81] , [i] Ref.[82], [j] Ref.[77], [k] Ref.[83], [l] Ref.[84] , [m] Ref.[75].

**Table 4.** Lattice constants (*a*), bulk moduli (*B*), shear moduli (*G*), and Young's moduli (*E*) for semiconductor structures. The Si and Ge systems are calculated using the diamond structure. The InTe system adopts the B1 structure. Other semiconductors are calculated using the zinc blende structure. The ABACUS package with numerical atomic orbitals (NAO) is used for DFT calculations with the PBE functional. The values in parentheses are the deviations of the DP and DPA-Semi model's predictions from the reference DFT computed values. We do not include *G* and *E* values here because they are often not discussed in most of the semiconductor experimental literature.

| Systems | Methods | *a* (Å) | *B* (GPa) | *G* (GPa) | *E* (GPa) |
|---|---|---|---|---|---|
| Si | DFT | 5.480 | 84 | 63 | 152 |
| | DP | 5.482 | 76 (-8) | 64 (1) | 150 (-2) |
| | DPA-Semi | 5.484 | 82 (-2) | 61 (-2) | 145 (-7) |
| | Exp. | 5.430[a] | 99[a] | | |
| Ge | DFT | 5.779 | 57 | 42 | 103 |
| | DP | 5.774 | 60 (3) | 43 (1) | 104 (1) |
| | DPA-Semi | 5.781 | 55 (-2) | 43 (1) | 103 (0) |
| | Exp. | 5.652[b] | 77[c] | | |
| SiC | DFT | 4.392 | 210 | 194 | 445 |
| | DP | 4.392 | 212 (2) | 195 (1) | 448 (3) |
| | DPA-Semi | 4.391 | 210 (0) | 194 (0) | 445 (0) |
| | Exp. | 4.360[d] | 225[e] | | |
| BAs | DFT | 4.819 | 129 | 127 | 288 |
| | DP | 4.816 | 131 (2) | 128 (1) | 291 (3) |
| | DPA-Semi | 4.819 | 130 (1) | 134 (7) | 300 (12) |
| | Exp. | 4.777[f] | 148[f] | | |
| BN | DFT | 3.621 | 370 | 392 | 870 |
| | DP | 3.621 | 369 (-1) | 401 (9) | 883 (13) |
| | DPA-Semi | 3.621 | 402 (32) | 389 (-3) | 882 (12) |
| | Exp. | 3.615[g] | 369[g] | | |
| AlN | DFT | 4.413 | 185 | 131 | 318 |
| | DP | 4.413 | 189 (4) | 129 (-2) | 317 (-1) |
| | DPA-Semi | 4.410 | 200 (15) | 129 (-2) | 319 (1) |
| | Exp. | 4.380[h] | 202[i] | | |
| AlP | DFT | 5.514 | 80 | 50 | 124 |
| | DP | 5.514 | 79 (-1) | 50 (0) | 124 (0) |
| | DPA-Semi | 5.513 | 79 (-1) | 50 (0) | 124 (0) |
| | Exp. | 5.470[j] | 86[j] | | |
| AlAs | DFT | 5.742 | 65 | 42 | 105 |
| | DP | 5.743 | 65 (0) | 41 (-1) | 101 (-4) |
| | DPA-Semi | 5.744 | 65 (0) | 43 (1) | 106 (1) |
| | Exp. | 5.660[j] | 82[j] | | |
| InP | DFT | 5.970 | 58 | 33 | 84 |
| | DP | 5.969 | 58 (0) | 31 (-2) | 79 (-5) |
| | DPA-Semi | 5.973 | 58 (0) | 33 (0) | 85 (1) |
| | Exp. | 5.868[r] | 71[s] | | |
| InAs | DFT | 6.200 | 48 | 26 | 67 |
| | DP | 6.201 | 49 (1) | 27 (1) | 68 (1) |
| | DPA-Semi | 6.202 | 48 (0) | 26 (0) | 67 (0) |
| | Exp. | 6.058[t] | 48[u] | | |
| InSb | DFT | 6.631 | 37 | 20 | 52 |
| | DP | 6.631 | 37 (0) | 22 (2) | 56 (4) |
| | DPA-Semi | 6.628 | 38 (1) | 18 (-2) | 47 (-5) |
| | Exp. | 6.473[v] | 46[w] | | |
| GaN | DFT | 4.555 | 170 | 112 | 276 |
| | DP | 4.555 | 171 (1) | 113 (1) | 278 (2) |
| | DPA-Semi | 4.554 | 156 (-14) | 111 (-1) | 267 (-9) |
| | Exp. | 4.52[h] | 190[i] | | |
| GaP | DFT | 5.517 | 77 | 54 | 132 |
| | DP | 5.518 | 76 (-1) | 53 (-1) | 130 (-2) |
| | DPA-Semi | 5.517 | 76 (-1) | 53 (-1) | 128 (-4) |
| | Exp. | 5.450[k] | 88[l] | | |
| GaAs | DFT | 5.750 | 61 | 43 | 106 |
| | DP | 5.752 | 61 (0) | 43 (0) | 105 (-1) |
| | DPA-Semi | 5.751 | 61 (0) | 39 (-4) | 97 (-9) |
| | Exp. | 5.650[m] | 76[m] | | |
| CdTe | DFT | 6.628 | 35 | 15 | 40 |
| | DP | 6.630 | 35 (0) | 15 (0) | 39 (-1) |
| | DPA-Semi | 6.630 | 35 (0) | 14 (-1) | 38 (-2) |
| | Exp. | 6.480[n] | 39[n] | | |
| InTe | DFT | 6.288 | 38 | 19 | 50 |
| | DP | 6.289 | 39 (1) | 19 (0) | 49 (-1) |
| | DPA-Semi | 6.288 | 39 (1) | 19 (0) | 50 (0) |
| | Exp. | 6.160[o] | 69[p] | | |
| CdSe | DFT | 6.213 | 45 | 18 | 47 |
| | DP | 6.213 | 44 (-1) | 16 (-2) | 42 (-5) |
| | DPA-Semi | 6.215 | 46 (1) | 17 (-1) | 46 (-1) |
| | Exp. | 6.050[n] | 53[n] | | |
| ZnS | DFT | 5.453 | 69 | 37 | 95 |
| | DP | 5.453 | 69 (0) | 35 (-2) | 89 (-6) |
| | DPA-Semi | 5.453 | 65 (-4) | 37 (0) | 94 (-1) |
| | Exp. | 5.410[q] | 77[q] | | |
| CdS | DFT | 5.939 | 53 | 20 | 55 |
| | DP | 5.939 | 52 (-1) | 18 (-2) | 49 (-6) |
| | DPA-Semi | 5.943 | 52 (-1) | 20 (0) | 56 (1) |
| | Exp. | 5.820[n] | 62[n] | | |

[a] Ref.[85], [b] Ref.[86], [c] Ref.[87], [d] Ref.[74], [e] Ref.[80], [f] Ref.[88], [g] Ref.[89], [h] Ref.[90], [i] Ref.[91], [j] Ref.[92], [k] Ref.[93], [l] Ref.[94],b[m] Ref.[95], [n] Ref.[96], [o] Ref.[97], [p] Ref.[98], [q] Ref.[99], [r] Ref.[100], [s] Ref.[101], [t] Ref.[102], [u] Ref.[103], [v] Ref.[104], [w] Ref.[105].

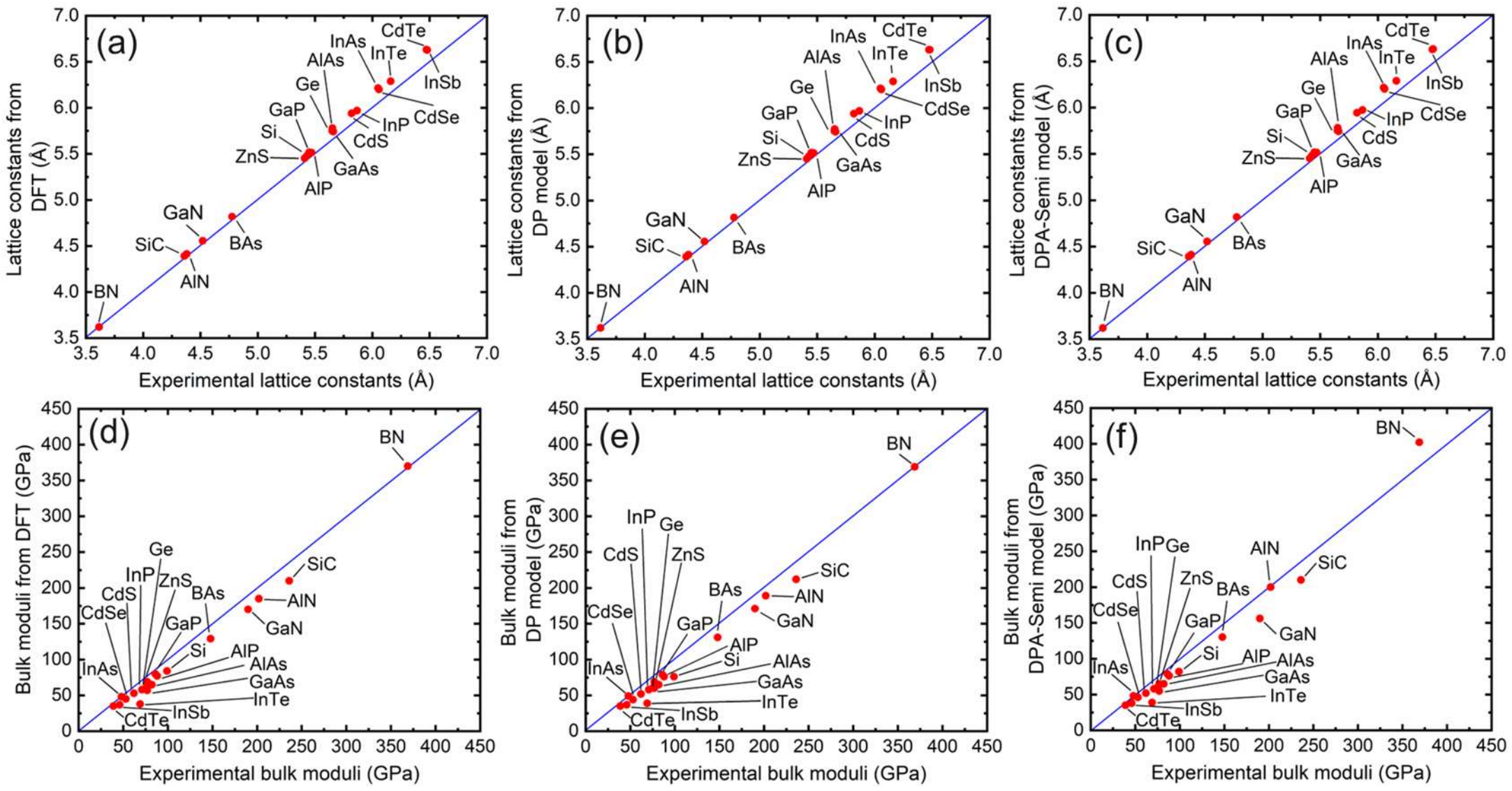

**Fig.3** Predicted lattice constants of various semiconductors by (a) DFT calculations with NAO basis set, (b) the DP models, (c) the DPA-Semi model, and available experimental data. Predicted bulk moduli of various semiconductors by (d) DFT calculations with NAO basis set, (e) the DP models, (f) the DPA-Semi model, and available experimental data. The DFT results are calculated by the ABACUS package with numerical atomic orbitals.

### 3.2 Melting temperatures

Estimating the melting temperatures of semiconductors can help one to study the solid-liquid phase transition and related thermodynamic properties. Each semiconductor has a unique melting temperature, and in order to predict it, we used the 'heat-until-melts' method for various semiconductors and the two-phase coexistence method for Si and SiC. The 'heat-until-melts' method was known to overestimate the melting temperatures due to the superheating effects[106, 107]. However, it would be useful here to yield upper boundaries of the melting temperatures for the semiconductors by using machine-learning-based models. We then performed MD simulations using the DP and DPA-Semi models to predict the melting temperatures and compared them with the experimental data. On the other hand, the two-phase coexistence method yields a more accurate melting temperature but the method is more expensive, and we adopted the DPA-Semi model for these simulations.

The initial structure setups for using the 'heat-unitl-melts' method are shown in Table 5. The number of atoms in these structures is selected to be at least 1000 atoms in order to eliminate the size effects. All of the crystal structures were simulated at an external pressure of 1 bar for 50 ps using the isothermal-isobaric (NPT) ensemble with a time step of 2 fs. The melting temperature is obtained by changing the simulation temperatures until the liquid phase of the semiconductor is observed by the mean square displacement (MSD). Fig. 4(a) compares the melting temperatures of semiconductors as obtained from the DP and DPA-Semi models. The results indicated that the DPA-Semi model well reproduces the results of DP potentials. Fig. 4(b) further compares the melting temperatures predicted by the DPA-Semi model and those from experimental results[108]. In general, the melting temperatures predicted by the DPA-Semi model are in reasonable agreement with the experimental values, while most predicted melting points are higher than the experimental values. For example, the predicted melting points of Si and SiC (3C) are about 2162 and 4337 K, higher than the experimental melting points of 1687 and 3100 K [108], respectively.

**Table 5.** Initial structures used for melting temperatures calculations of semiconductors. Labels α, β, and γ are the angles (degree) between the crystallographic axes of a crystal, respectively. The number of atoms in the cell is also listed.

| Systems | Crystal Structures | α | β | γ | Number of Atoms |
|---|---|---|---|---|---|
| Si | Cubic ($Fd\bar{3}m$) | 90 | 90 | 90 | 1000 |
| Ge | Cubic ($Fd\bar{3}m$) | 90 | 90 | 90 | 1000 |
| SiC | Cubic ($F\bar{4}3m$) | 90 | 90 | 90 | 1000 |
| BAs | Cubic ($F\bar{4}3m$) | 90 | 90 | 90 | 1000 |
| AlN | Hexagonal ($P6_3/mmc$) | 90 | 90 | 120 | 1536 |
| AlP | Cubic ($F\bar{4}3m$) | 90 | 90 | 90 | 1000 |
| AlAs | Cubic ($F\bar{4}3m$) | 90 | 90 | 90 | 1000 |
| InP | Cubic ($F\bar{4}3m$) | 90 | 90 | 90 | 1000 |
| InAs | Cubic ($F\bar{4}3m$) | 90 | 90 | 90 | 1000 |
| InSb | Cubic ($F\bar{4}3m$) | 90 | 90 | 90 | 1000 |
| GaP | Cubic ($F\bar{4}3m$) | 90 | 90 | 90 | 1000 |
| GaAs | Cubic ($F\bar{4}3m$) | 90 | 90 | 90 | 1000 |
| CdTe | Cubic ($F\bar{4}3m$) | 90 | 90 | 90 | 1000 |
| InTe | Cubic ($Fm\bar{3}m$) | 90 | 90 | 90 | 1000 |
| CdSe | Cubic ($F\bar{4}3m$) | 90 | 90 | 90 | 1000 |
| ZnS | Cubic ($F\bar{4}3m$) | 90 | 90 | 90 | 1000 |
| CdS | Cubic ($F\bar{4}3m$) | 90 | 90 | 90 | 1000 |

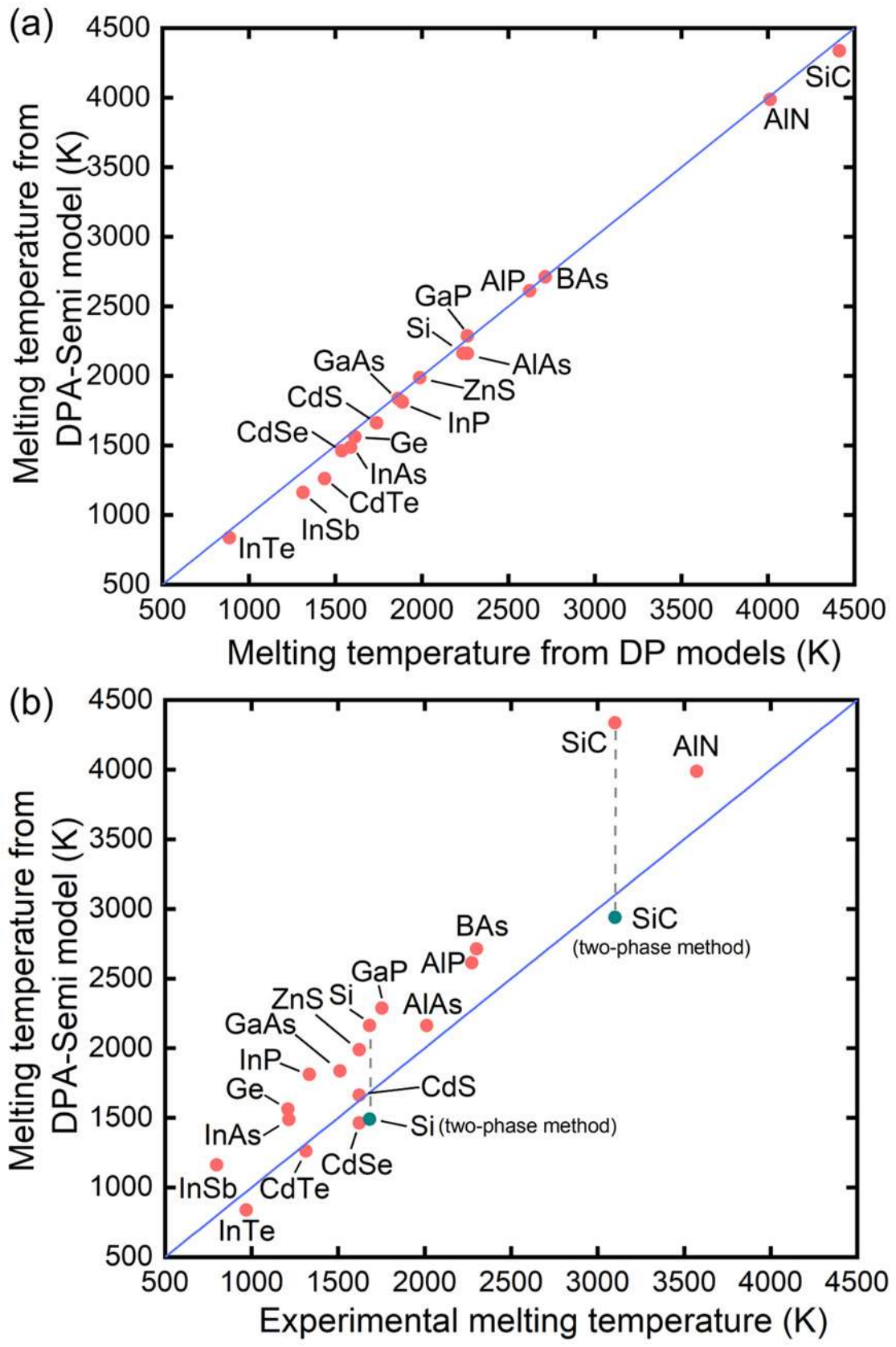

**Fig.4** (a) Predicted melting temperatures of various semiconductors by the DPA-Semi and DP models. (b) Predicted melting temperatures of various semiconductors by the DPA-Semi and available experimental data.[108] In addition, the melting points of Si (diamond structure) and SiC (3C structure) obtained from the two-phase coexistence method are shown for comparision (green dots).

Next, we utilized the two-phase coexistence method[109] to predict a more accurate melting temperature of Si and SiC (3C) by using the DPA-Semi model. The method predicts the coexistence temperature of solid and liquid phases and can be used to obtain a more accurate melting temperature.[107] We initially set up a 2048-atom solid phase structure and a 2048-atom liquid phase for both Si and 3C-SiC structures. The solid phase was heated to 1000 K, while the liquid phase was heated slightly above the melting temperatures predicted through the abovementioned 'heat-unitl-melts' method. The simulations were run for 30 ps with the time step being 1 fs to reach equilibrium structures for both phases. We then combined the two phases to construct a 4096-atom cell and fixed the solid-phase atoms. Subsequently, a 30 ps simulation was run using the NVT ensemble at the experimentally determined melting temperature to reduce the interfacial energy between the solid and liquid phases. Finally, we conducted an NPT simulation of the entire system at a pressure of 1 bar. We incrementally increased the temperature with an interval of 5 K and performed simulations of 200 ps to determine the coexistence temperature of the solid and liquid phases. This coexistence temperature served as the melting temperatures obtained through the two-phase method.

Figs. 5(a) and (b) show the initial structure and a structure of 200 ps for the solid-liquid coexistence DPMD simulations of Si and SiC (3C), respectively. In the case of Si, the DPA-Semi model yields a coexistence temperature close to 1485 K, which is also shown in Fig. 4(b). When compared to the experimental data of 1687 K[108], the coexistence method yields a better melting temperature than the 'heat-until-melts' method. This result is also consistent with 1485 K from a recent DPMD simulation using the two-phase coexistence method.[110] On the other hand, the DPA-Semi model for 3C-SiC yields a solid-liquid coexistence temperature of ~2925 K, as shown in Fig. 4(b). Compared with the overheating effect of the "heat-until-melts" method, the data is closer to the experimental melting temperature of 3100 K[108].

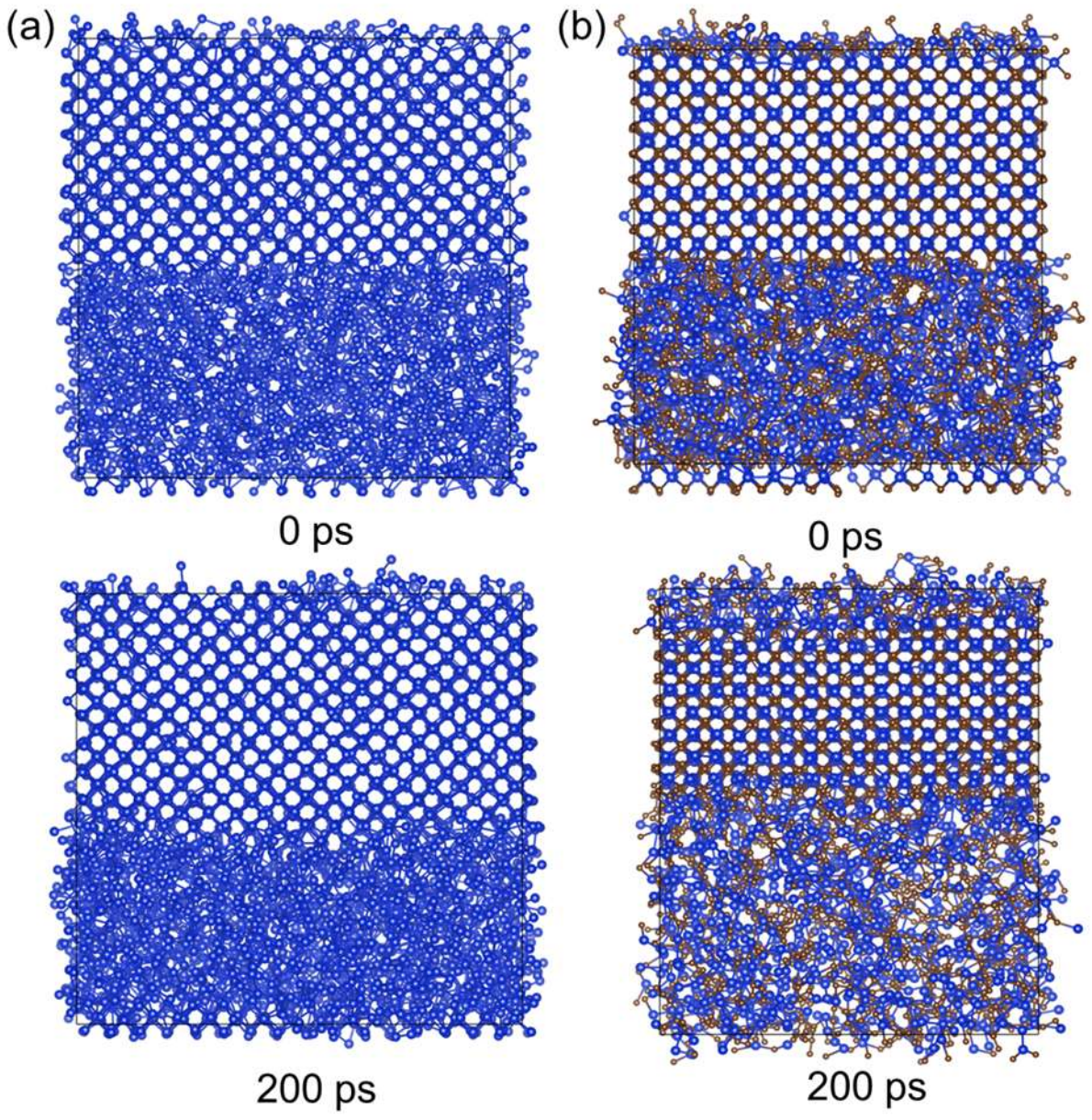

**Fig.5** Two-phase coexistence method with the DPA-Semi model to predict the melting points of Si and SiC (3C) structures. (a) At 1485 K, the Si system with 4096 atoms (blue) exhibit coexistence of solid and liquid phases during a 200 ps DPMD simulations. (b) At 2925 K, the 3C-SiC system with 4096 atoms (Si: blue, C: brown) show solid-liquid coexistence during a 200 ps DPMD.

Note that we do not include the melting temperatures of GaN and BN in the melting temperature simulations because the experimental results for the two structures are still inconclusive. One the one hand, the phase diagram of GaN has not yet been well understood through research because significant phase transformations occur[111]. For instance, the liquid GaN may coexist with $N_2$ molecules, which poses challenges to the model. Currently, both DP and DP-Semi models do not include training sets of $N_2$ molecules. However, it is possible to fine-tune the DP-Semi model with additional data of $N_2$ molecules for downstream tasks but beyond the scope of current work.

On the other hand, measuring the melting point of BN remains a significant challenge due to the difficulties associated with heating optically transparent diamonds through the absorption of intense laser radiation in a diamond anvil cell, as well as the practical obstacles in attaining high temperatures within a large-volume press using conventional resistance heating techniques.[112] Besides, BN easily sublimes at its melting temperature[113]. Due to the challenges in simulating the liquid phases of GaN and BN, we do not include the predicted melting temperatures of GaN and BN in this work.

### 3.3 Phonon spectra

The phonon spectrum provides crucial insights into the vibrational properties of a material, shedding light on its thermal, electrical, and mechanical behavior. Furthermore, the phonon band structure offers valuable information on the dispersion relation of phonon modes. In this regard, Fig.6 shows the computed phonon spectra of 19 semiconductors utilizing the DP, DPA-Semi, and DFT calculations based on the Phonopy 2.20.0 package[114, 115]. For both DP and DPA-Semi calculations, a 5×5×5 supercell was employed to ensure convergence, while for DFT-based computations, we utilized a 4×4×4 supercell structure that is large enough to converge the phonon spectra. We find the phonon spectra obtained from the three methods for AlAs, AlN, AlP, GaN, GaAs, GaP, ZnS, CdS, CdSe, CdTe, InP, InAs, Bas, BN, and SiC agree well. However, some deviations still exist for the Si, Ge, InSb, and InTe structures. In general, we consider the DPA-Semi model is able to yield satisfactory results for the phonon spectra of semiconductors.

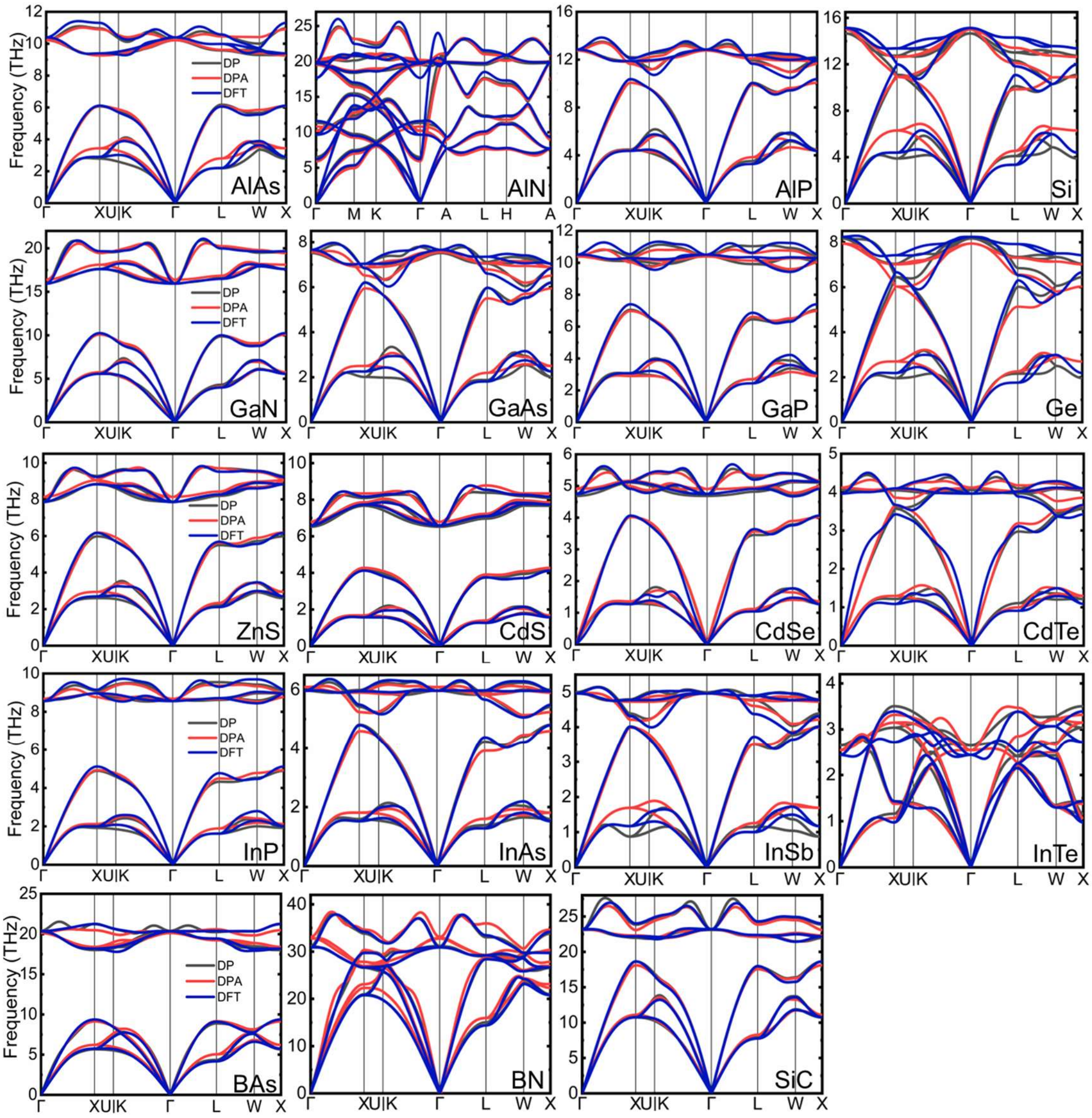

**Fig.6** Phonon spectra of 19 semiconductors calculated by the DFT, DPA-Semi, and DP methods. The Si and Ge systems adopt the diamond structure while the AlN system adopts the B1 structure; other semiconductors are calculated using the zinc blende structure. The special k-points are Γ(0,0,0), M(0.5,0,0), K(1/3, 1/3,0), A(0,0,0.5), L(0.5,0,0.5), H(1/3, 1/3,0.5) for the B1 structure, and Γ(0,0,0), X(0.5,0,0.5), U(0.625,0.25,0.625), K(0.375 0.375 0.75), L(0.5,0.5,0.5), W(0.5,0.25,0.75) for the diamond and zinc blende structures.

### 3.4 Liquid and Amorphous Structures

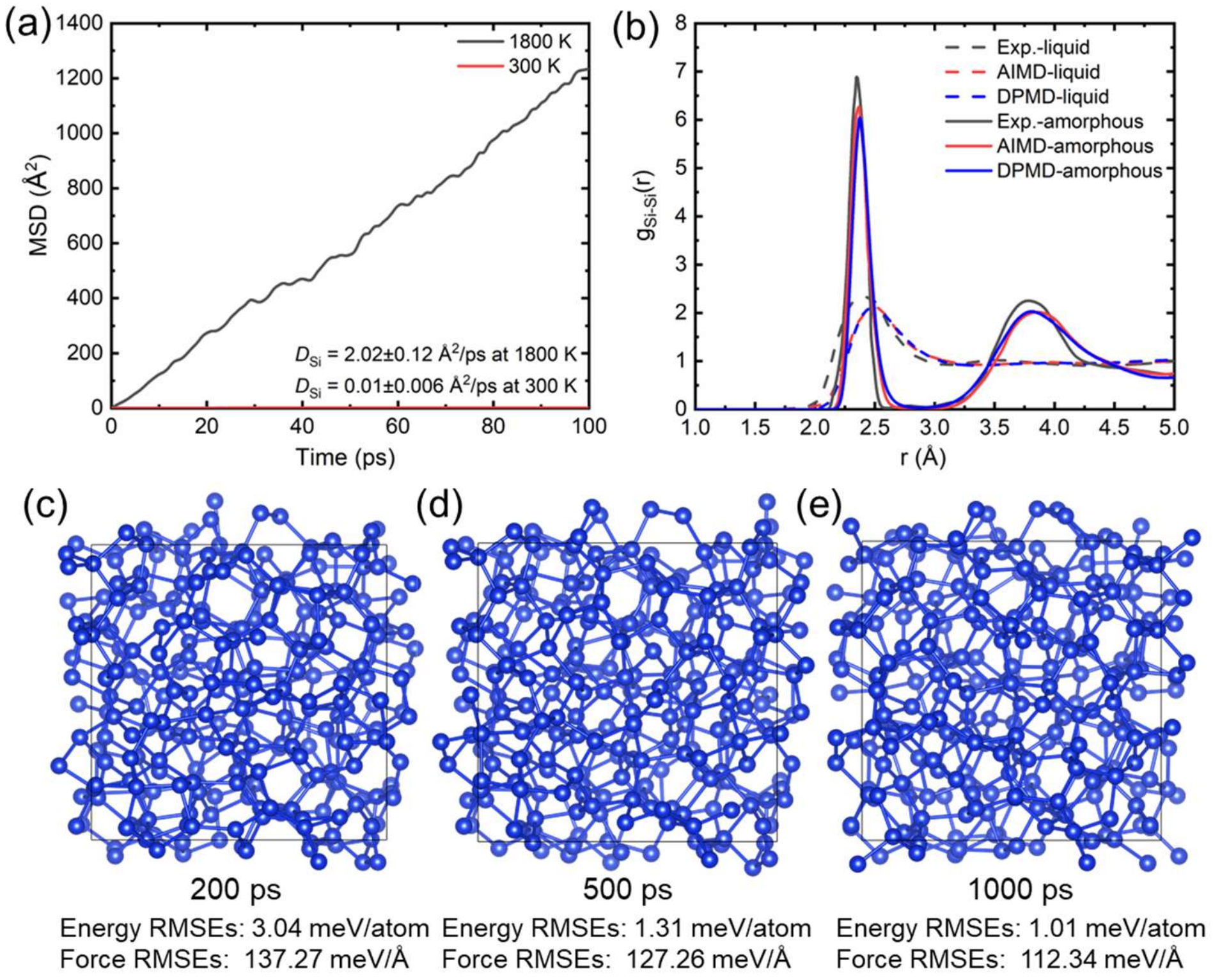

**Fig.7** (a) Mean square displacements (MSD) and diffusion coefficients (*D*) of crystalline Si at 300 K (red line) and liquid Si at 1800 K (black line) from the DPMD trajectories with the DPA-Semi model. (b) Radial distribution functions g(r) of Si-Si for liquid Si at 1800 K and amorphous Si at 300 K, where the black dashed line indicates the experiment results,[116, 117] the red dashed line depicts AIMD (PBE) results,[38, 118] and the blue dashed line represents the DPMD results with the DPA-Semi model. (c), (d) and (e) show the snapshots of amorphous Si at 300 K when the simulation time is 200, 500, and 1000 ps, respectively. RMSEs of the total energy (meV/atom) and forces (meV/Å) for the snapshots are also shown.

The DPA-Semi model was adopted to predict properties of various phases of Si.[38, 119, 120, 121, 122] We first prepared a crystalline Si with a 3×3×3 supercell and 216 Si atoms arranged in a diamond lattice configuration. The system was first heated from 300 to 2,500 K for 1,000 ps to yield a liquid structure, then simulated at 2,500 K for another 1,000 ps to ensure an equilibration state within the isothermal-isobaric (*NPT*) ensemble and a time step of 1.0 fs. A controlled cooling period was subsequently initiated, gradually lowering the temperature to 1,800 K throughout at least 500 ps to achieve the liquid phase and to 300 K over a minimum of 1500 ps to obtain the amorphous phase. Equilibration runs were performed in the *NPT* ensemble at 0 bar and 1,800 K/300 K for at least 1,000 ps to obtain properties of the liquid/amorphous phases of Si.

The mean square displacements (MSD) of liquid and crystalline phases of Si are shown in Fig. 6(a). Here, we conducted a 1000 ps NPT simulation on crystalline Si at 300 K. By using the formula $\frac{1}{6}\lim_{\Delta t\to\infty}\frac{MSD(\Delta t)}{\Delta t}$, the diffusion coefficient ($D$) of crystalline Si at 300 K and liquid Si at 1800K are computed to be 0.01±0.006 Å$^2$/ps and 2.02±0.12 Å$^2$/ps.

Fig. 7(b) compares the radial distribution functions g(r) of Si-Si for liquid Si at 1,800 K and amorphous Si at 300 K. We find that the first peak position of the $g_{Si-Si}(r)$ of DPMD with the DPA-Semi model for liquid phase or amorphous phase agrees well with previous AIMD results with the PBE functional[38, 118]. To further evaluate the accuracy of the DPA-Semi model for amorphous Si, Figs.7(c), (d), and (e) show a few snapshots from the amorphous Si trajectory, and we compared the energy and force obtained from the DPA-Semi model to those obtained from the DFT method. The deviations in energy and force are relatively small, demonstrating the DPA-Semi owns an accuracy similar to the DFT method in describing the amorphous structures of Si. However, it is worth noting that simulation results exhibit some deviations when compared to the experimental data[116, 117], as the peak positions from simulations are slightly larger for the liquid phase of Si[116], while the peak heights from simulations are slightly lower than the experimental value for the amorphous phase of Si[117]. The above deviations have been attributed to the inaccuracy of PBE in simultaneously capturing covalent and metallic bonding in Si.[118,78]

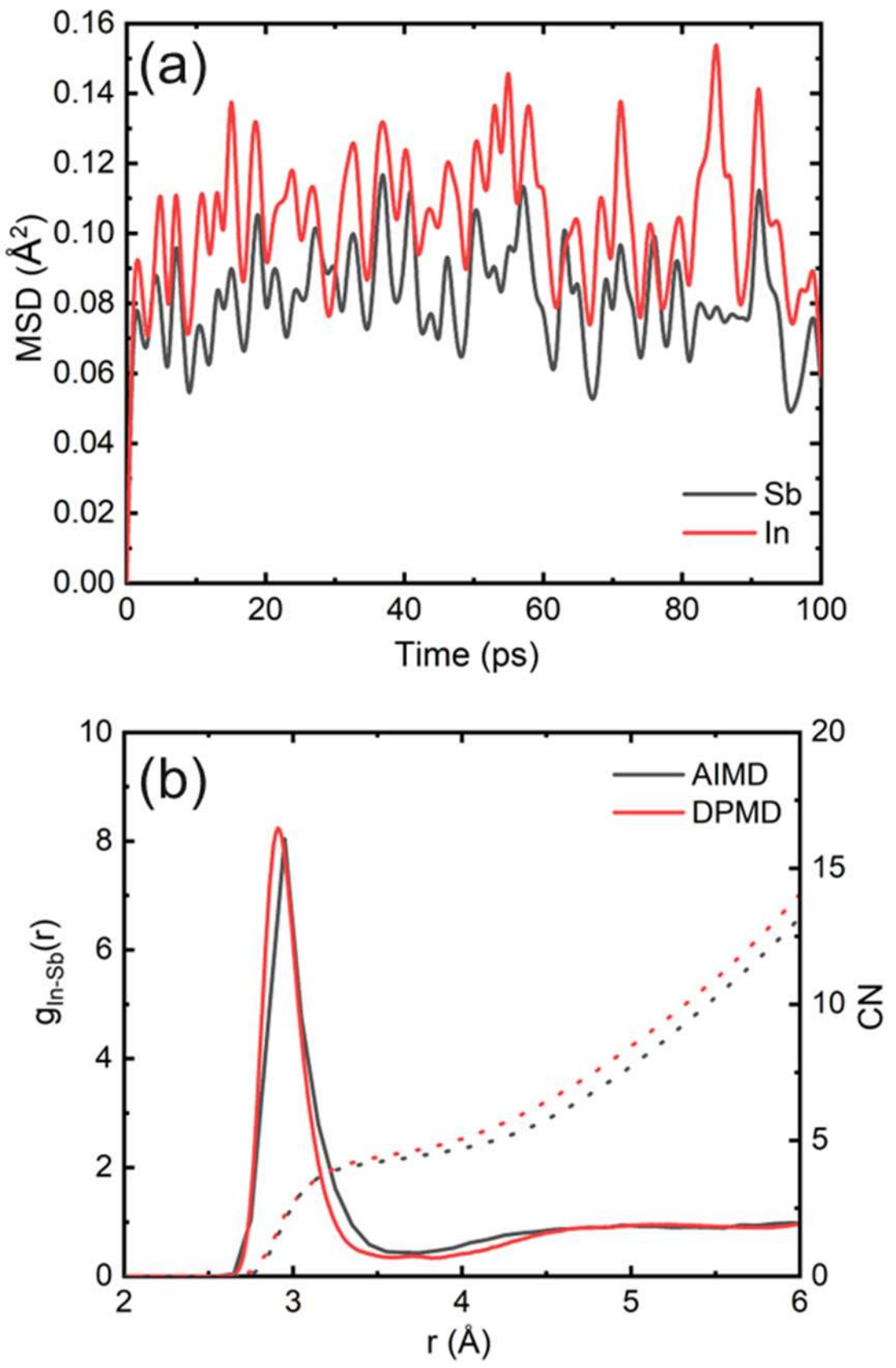

**Fig. 8** (a) Mean square displacements (MSD) for In and Sb atom at 300 K based on the DPMD simulations with the DPA-Semi model. (b) Radial distribution functions g(r) (solid lines) and coordination numbers (CN, dashed lines) for amorphous InSb at 300 K. The black line indicates our AIMD results (64 atoms) using the PBE functional, the red line denotes the DPMD with the DPA-Semi model results (216 atoms).

We also prepared an amorphous phase of InSb at a density of 6.09 g/cm$^3$ containing 64/216 atoms, which used the experimental density (6.1 g/cm$^3$) of amorphous InSb measured by X-ray reflectivity[123]. The method to generate the amorphous structure of InSb is similar to the abovementioned amorphous Si. Concurrently, a 10-ps (64 atoms) and 1,000-ps (216 atoms) DPMD trajectory were performed with the DFT method and the DPA-Semi model. The canonical ensemble (*NVT*) was selected with a time step of 1.0 fs at a temperature of 300 K.

Similarly, we also calculated the diffusion coefficients (*D*) of In and Sb atoms. The MSD is shown in Fig. 8(a), and the *D* is close to zero for In and Sb atoms, indicating that the In and Sb atoms are in an amorphous phase structure. Radial distribution functions and coordination number (CN) for the two methods are compared in Fig. 8(b). We observed the CN of In-Sb at 300 K is about 4.9. Although the $g_{In\text{-}Sb}(r)$ of DPMD with the DPA-Semi model agrees well with that of AIMD, those are also shifted to slightly smaller distances due to DPA-Semi model bias. The positions of the first peak of $g_{In\text{-}Sb}(r)$ are at 2.96 Å from AIMD and 2.91 Å from DPMD with the DPA-Semi model, respectively. Those results are slightly larger than the experimental value (2.82 Å) obtained from x-ray measurement data.[123] We utilized the DPA-Semi model to assess the RMSEs of energy and force for various configurations in AIMD, revealing a non-negligible discrepancy (the energy and force RMSEs are 58.3 meV/atom and 201.8 meV/Å, respectively). This may explain the deviation between the RDF peak values obtained from DPA-Semi model calculations and those from DFT calculations.

## 4. Conclusions

In this work, we generated an attention-based deep potential model (i.e., DPA-Semi model) and a series of DP models that allow for large-scale pretraining on atomistic datasets to study the various properties of 19 semiconductors ranging from group IIB to VIA, namely, Si, Ge, SiC, BAs, BN, AlN, AlP, AlAs, InP, InAs, InSb, GaN, GaP, GaAs, CdTe, InTe, CdSe, ZnS, CdS. Importantly, the DPA-Semi model, along with datasets, exhibits a high degree of generalization capability for downstream tasks. Researchers can adopt our semiconductor datasets by augmenting them with a small amount of new data to effectively simulate defect and dopant-incorporated semiconductors using the DPA method, thereby reducing the amount of data required in training the traditional DP model.

We focused on comparing the DP and DPA-Semi model results. For example, we calculated the lattice constants and elastic moduli for 19 semiconductors ranging from group IIB to VIA. Our results indicated that the DPA-Semi model can well reproduce the results of each DP model and are in excellent agreement with the DFT results calculated by the ABACUS package. Besides, we also adopted the 'heat-unti-melts' method and the 'two-phase coexistence' method to predict the melting temperatures of semiconductors using the DP or DPA-Semi models. The results indicated that the DPA-Semi model can accurately reproduce the results of each DP model. The melting temperatures predicted by the DPA-Semi model are close to the experimental values.

We took SiC as an example and found the energy differences of the 3C-, 2H-, 4H-, and 6H-SiC polytypes calculated by DFT and DPA-Semi model were small, on the order of meV/atom. Besides, the total energies per atom satisfied the inequality relation: $E_{2H} > E_{3C} > E_{6H} > E_{4H}$, which agreed with the previous theoretical results. It indicated that our DFT calculations based on ABACUS with the LCAO method and DPA-Semi calculation results could correctly distinguish these structural energy differences. Besides, we also compared the lattice constants and elastic constants of the 3C-, 2H-, 4H-, and 6H-SiC structures calculated by ABACUS, DP model, and DPA-Semi model. The results also agreed well with the results from other packages (CASTEP, Quantum Espresso, ABINIT[72], e.g.) and experimental data, demonstrating that the DP and DPA-Semi models were adequate for studying the mechanical properties of SiC polytypes. In addition, we found that the radial distribution functions of Si-Si for liquid Si and radial distribution functions of In-Sb for the amorphous phase of InSb calculated by DPMD with the DPA-Semi model were in reasonable agreement with those from AIMD.

In conclusion, our work provided reliable evidence that the DPA-Semi model can be readily employed to study the scientific issues of group IIB to VIA semiconductor systems with GGA quality accuracy. In the future, the DPA-Semi model can be potentially useful for various downstream tasks and substantially reduce computational costs.

## Author Contributions

† These authors contributed equally to this work.

## Conflicts of interest

There are no conflicts of interest to declare.

## Acknowledgments

This work is supported by the National Science Foundation of China under Grant No.12122401 and No.12074007. The numerical simulations were performed on the high-performance computing platform of CAPT and the "Bohrium" cloud computing platform of DP Technology Co., LTD.